\documentclass{PoS}

\newcommand{\msun}{\ensuremath{M_{\odot}}}

\newcommand{\mdot}{\ensuremath{\dot{M}}}

\newcommand{\beq}{\begin{equation}}
\newcommand{\eeq}{\end{equation}}
\newcommand{\beqa}{\begin{eqnarray}}
\newcommand{\eeqa}{\end{eqnarray}}

\newcommand{\nbeq}{\begin{equation*}}
\newcommand{\neeq}{\end{equation*}}

\newcommand{\xmm}{\emph{XMM-Newton}}
\newcommand{\inte}{\emph{INTEGRAL}}
\newcommand{\swift}{\emph{Swift}}
\newcommand{\chandra}{\emph{Chandra}}

\newcommand{\lsim}{\raisebox{-.4ex}{$\stackrel{<}{\scriptstyle \sim}$}}
\newcommand{\msim}{\raisebox{-.4ex}{$\stackrel{>}{\scriptstyle \sim}$}}

\newcommand{\aj}{AJ}%
\newcommand{\apj}{ApJ}%
\newcommand{\apjl}{ApJ}%
\newcommand{\apjs}{ApJS}%
\newcommand{\aap}{A\&A}%
\newcommand{\aapr}{A\&A~Rev.}%
\newcommand{\ATel}{ATel}%
\newcommand{\mnras}{MNRAS}%
\title{Supergiant Fast X-ray Transients - A short review}

\ShortTitle{Supergiant Fast X-ray Transients - A short review}

\author{\speaker{Lara Sidoli}%
         \thanks{A footnote may follow.}\\
        INAF/IASF Milano Italy\\
        E-mail: \email{sidoli@iasf-milano.inaf.it}}

\abstract{
I present a brief up-to-date review of the current understanding of Supergiant Fast X--ray Transients, 
with an emphasis on the observational point of view.
After more than a decade since their discovery, a remarkable progress has been made in getting
the picture of their phenomenology at X--ray energies. 
However, a similar in-depth investigation of the properties of the supergiant companions is needed, but   
has started more recently. 
A multifrequency approach is the key to fully understand the physical mechanism
driving the SFXT behaviour, still under debate.
}

\FullConference{XII Multifrequency Behaviour of High Energy Cosmic Sources Workshop\\
		 12-17 June 2017 \\
		 Palermo, Italy}

\begin{document}

\section{The strange case of SFXTs}

In this short review  paper I will concentrate on the most recent achievements about Supergiant Fast X--ray Transients (SFXTs),
emphasizing an observational standpoint. The X--ray window will be looked at in particular, but with an eye also to observations at 
other frequencies, especially for what concerns the future directions.
I will mainly focus on advances on the topic obtained thanks to systematic investigations of SFXT class as a whole,
while I refer the reader to \cite{Walter2015} for a summary of the properties of the single sources.
For a unique, comprehensive and detailed review of the current understanding of the interplay 
between the outflowing wind from massive stars and accretion processes in wind-fed supergiant HMXBs, 
I refer the reader to \cite{Martinez-Nunez2017}.

\begin{figure}[ht!]
\begin{center}
\hspace{-0.5cm}
\includegraphics[height=9.0cm, angle=0]{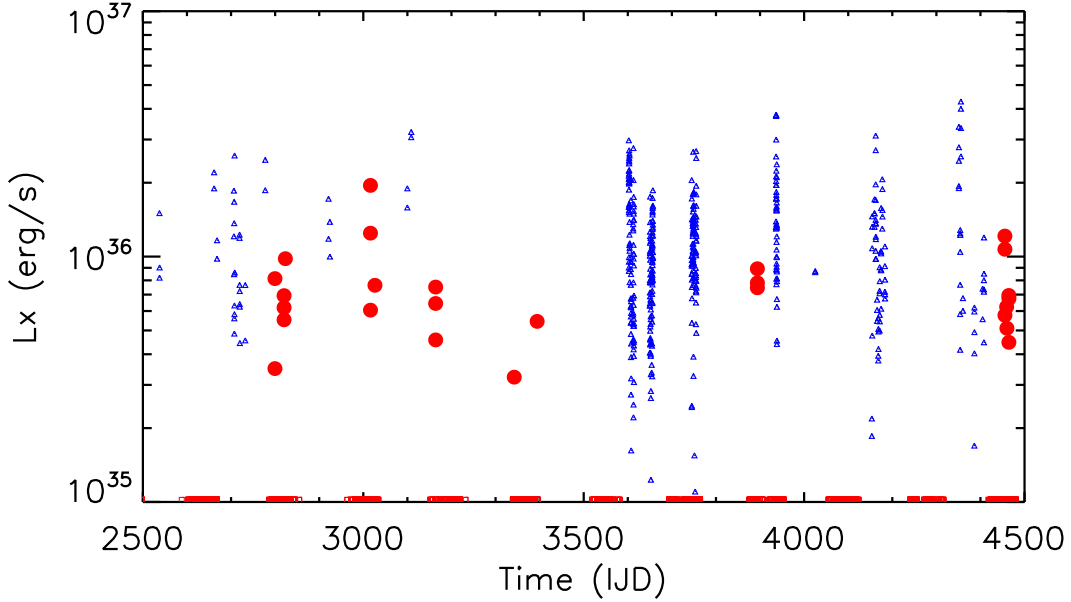} \\
\end{center}
\vspace{0.0cm}
\caption{\scriptsize 
Comparison between \inte/IBIS  light curves (bin time of 2~ks; 17--50 keV) of 
Vela X-1 (blue triangles) and an SFXT (IGR~J17544-2619; each red solid cirle mark a flare).
Red bars on the x-axis indicate times when IGR~J17544-2619 was undetected 
by \inte\ (its duty cycle is $\lsim$1~$\%$). On the other hand, Vela X--1 is always detected. 
Time is in units of \inte\ JD (IJD=MJD-51544). 
}
\label{lsfig:comp_lc}
\end{figure}

SFXTs (\cite{Sguera2005}, \cite{Negueruela2005a}) are a sub-class of High Mass X--ray Binaries (HMXBs) 
where a compact object (typically a neutron star) accretes
a fraction of the clumpy wind from the blue supergiant companion, triggering rare and short (less than a few days) outbursts, 
made of a number of bright X--ray flares (each flare with a duration of $\sim$thousands seconds). 
Most SFXTs were discovered by the \inte\ satellite (launched in 2002; \cite{Winkler2003}) as hard (above 17 keV) X--ray sources 
detected only during a short duration (a few hours) X--ray activity. 
Thanks to X--ray sky positions refined at arcsec level at soft X--rays (1-10 keV), 
they were quickly associated with O or B-type supergiants. 
The identification of these X--ray transients with massive X--ray binaries was somehow surprising, 
since the classical HMXBs with supergiant companion (SgHMXBs) known since the birth of X--ray astronomy 
are persistent X--ray emitters, with a limited range of intensity variability (around a factor of $\sim$10). 
Fig.~\ref{lsfig:comp_lc} shows the comparison between long-term X--ray \inte\ light curve of an SFXT and Vela X--1, the prototype of  persistent HMXBs.

Since their discovery, more than a decade ago, a huge observational effort has involved both 
long-term, X--ray monitoring investigations 
(summarized by \cite{Smith2012} using $RXTE$, 
by \cite{Paizis2014} and \cite{Sidoli2016} using \inte, 
by \cite{Bozzo2015} and \cite{Romano2015} using \swift)
and deep, very sensitive, X--ray pointings (by \chandra, e.g. \cite{zand2005}, and \xmm, 
starting from the pioneering observations discussed by \cite{Gonzalez2004}, 
to the more recent works by \cite{Drave2014}, \cite{Boon2016}, \cite{Sidoli2017}, 
ending with the latest published paper on SFXTs with \xmm\ \cite{Bozzo2017}). 

These X--ray observations led to the characterization of the SFXT properties, as follows:

\begin{enumerate}

\item {\em low} duty cycle ($\lsim5\%$) in bright X--ray flares (at L$_{X}$ $\msim$ 10$^{36}$~erg~s$^{-1}$) 

\item {\em high} dynamic range (ratio between maximum and minimum X--ray luminosity) $\msim$100 

\item {\em low} time-averaged luminosity L$_{X}$ $\lsim$ 10$^{34}$~erg~s$^{-1}$ 

\end{enumerate}

Within the above limits, a variety of behaviours is present among the members of the class. 
The duty cycle (percentage of time spent in bright X--ray activity) can reach very low values, as low as 0.1$\%$ (IGR~J08408-4503, \cite{Paizis2014}).
The dynamic range of their X--ray emission, between quiescence and flare peaks, can spand values between $\sim$100 (in this case the source
is named ``intermediate SFXT''), and 10$^6$, reached only in IGR~17544-2619, the prototypical SFXT where the X--ray luminosity variations are carried to their
extremities, from 10$^{32}$~erg~s$^{-1}$ (\cite{zand2005}, \cite{Drave2014}) 
to 10$^{38}$~erg~s$^{-1}$ \cite{Romano2015giant}.

Although it is evident from long-term \swift/XRT monitoring observations of a sample of SFXTs 
that the time-averaged X--ray luminosity is 
L$_{X}$ $\lsim$ 10$^{34}$~erg~s$^{-1}$
(\cite{Sidoli2008}, \cite{Bozzo2015} and \cite{Romano2015}), \swift/XRT is not sensitive enough to
reach very low fluxes and characterize the lowest luminosity states with short snapshots.
The percentage of time spent below the  \swift/XRT sensitivity threshold ($\sim$2$\times$10$^{-12}$~erg~~cm$^{-2}$~s$^{-1}$) 
ranges from 5\% to 67\%, depending on the source \cite{Romano2015}.
For instance,  there are three SFXTs (IGR~J08408--4503, IGR~J16328--4726 and IGR~J17544--2619)
that remained undetected by \swift/XRT most of the time (67$\%$, 61$\%$ and 55$\%$, respectively), 
with a 2--10 keV luminosity below 
2.6$\times$10$^{33}$~erg~s$^{-1}$ (IGR~J08408--4503),  
2.5$\times$10$^{34}$~erg~s$^{-1}$ (IGR~J16328--4726) and
2.1$\times$10$^{33}$~erg~s$^{-1}$ (IGR~J17544--2619).

Very low quiescent luminosities of L$_{X}$$\sim$10$^{32}$~erg~s$^{-1}$  
have been caught only in a few SFXTs  thanks to \xmm, \chandra, $Suzaku$ pointings  
(e.g. \cite{zand2005}, \cite{Bozzo2008_atel}, \cite{Drave2014}).
Remarkably, the lowest X--ray emission state ever reported in an SFXT was observed in  SAX~J1818.6--1703 using \xmm,
leading to a 3-$\sigma$ upper limit of F$_{X}$$<$1.1$\times$10$^{-13}$~erg~cm$^{-2}$~s$^{-1}$ 
(corrected for the absorption, 0.5--10 keV; \cite{Bozzo2008_atel}).
This implies an upper limit to its luminosity of L$_{X}$$<$6$\times$10$^{31}$~erg~s$^{-1}$ (at 2.1~kpc).
At present, it is unknown how much time each SFXT spends 
in quiescence at a luminosity level of 10$^{32}$~erg~s$^{-1}$. 
It is also possible that some more active (``intermediate'') SFXTs never reach it.
A monitoring programme with \xmm\ would be (unrealistically) needed, for this purpose.

The ample range of observed  behaviours has posed an issue about the possible in-homogeneity of the SFXT class 
and about the membership of some specific sources 
(e.g. the case of IGR~J16418-4532 or IGR~J16465-4507, e.g. \cite{Romano2015}, \cite{Walter2015} opposite to the 
classification reported by \cite{Paizis2014}).
IGR~J11215--5259 is also peculiar, since it is the only SFXT where X--ray outbursts 
are strictly periodic \cite{SidoliPM2006}, recurring every $\sim$165~days (\cite{Sidoli2007}, \cite{Romano2009}),
believed to be the orbital period of the system.
In general, also X--ray flares from other SFXTs appear more concentrated  
around the periastron passage (e.g. \cite{Gamen2015}), but nevertheless they can 
occur also at other orbital phases (\cite{Goossens2013}, \cite{Smith2012})
and remarkably, in all sources but IGR~J11215--5259, 
not every periastron passage triggers luminous flares. 
For example, after the recent determination of the orbital geometry in 
IGR~J08408-4503 (\cite{Gamen2015}), it is now evident that the long
X--ray observation performed with $Suzaku$ by \cite{Sidoli2010igr08408} 
covered a large orbital phase interval just around
periastron, and no bright flares were caught.
However, it is important to remark that in a few SFXTs the orbital period has been discovered 
thanks to a periodic modulation in their long-term X--ray light curve (sometimes still 
present after removing bright flares).

\begin{figure}[ht!]
\begin{center}
\hspace{-0.5cm}
\includegraphics[height=8.0cm, angle=0]{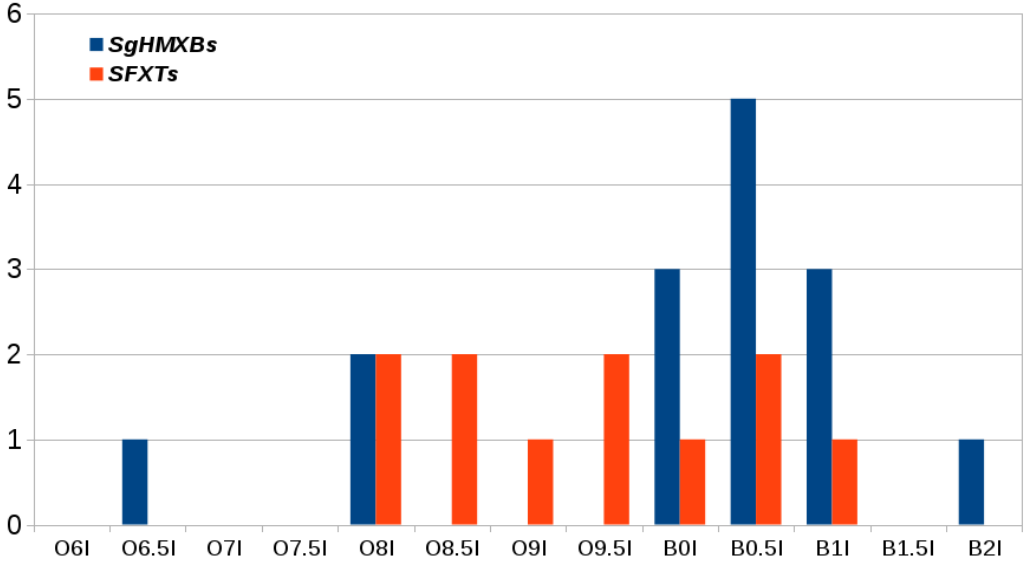} \\
\end{center}
\vspace{0.0cm}
\caption{\scriptsize  Spectral type of the supergiant companions in classical SgHMXBs (in {\em blue}, 
taken from the list reported by \cite{Martinez-Nunez2017}) 
and in SFXTs (in {\em orange}, taken from Table~\ref{lstab:sfxt}).
}
\label{lsfig:spetype}
\end{figure}

\begin{figure}[ht!]
\begin{center}
\hspace{-1.5cm}
\includegraphics[height=12.0cm, angle=0]{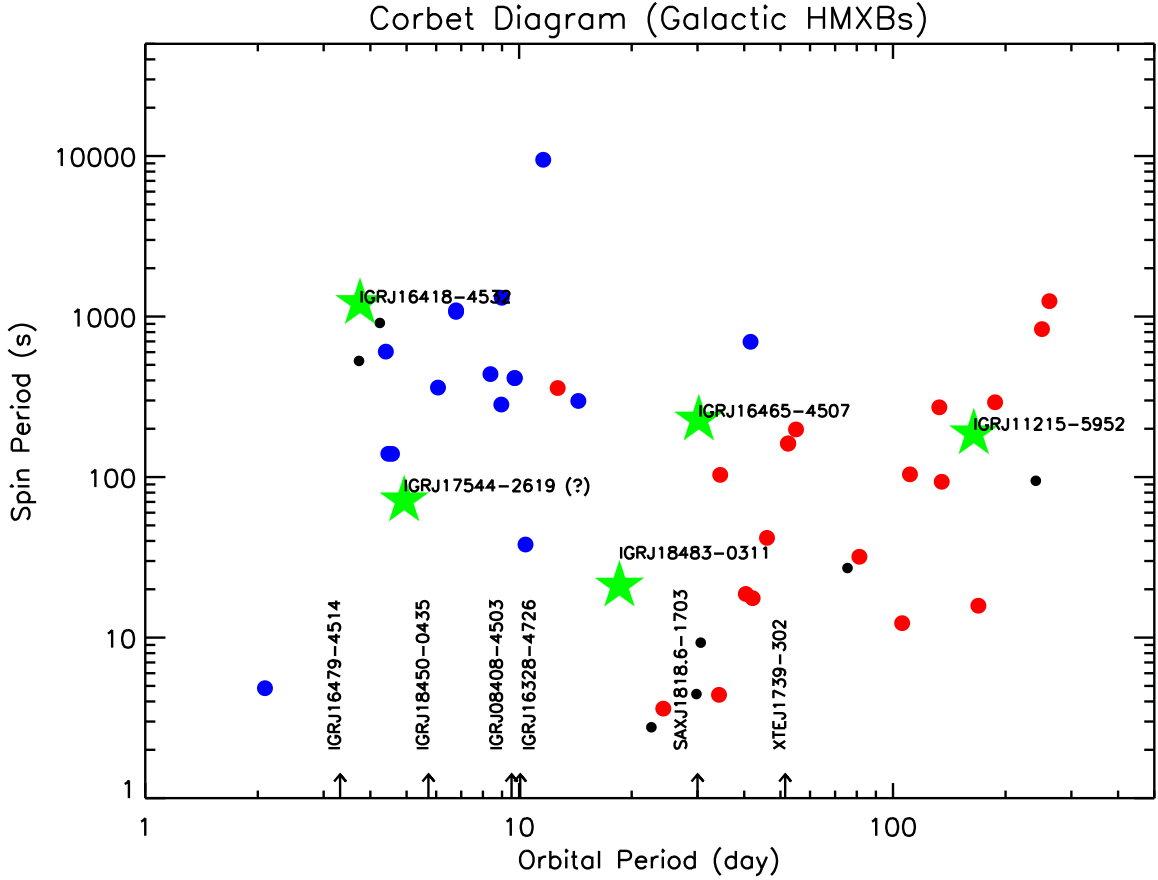} \\
\end{center}
\vspace{-0.5cm}
\caption{\scriptsize  
Spin period versus orbital period for Galactic High Mass X--ray Binaries.
{\em  Blue dots} indicate HMXBs hosting supergiant companions 
(excluding SFXTs, that are marked with  {\em green stars}), while the
{\em  red dots} mark Be/XRBs. Arrows on the x-axis mark the orbital 
periods of SFXTs with still unknown pulsar spin periods.
}
\label{lsfig:corbet}
\end{figure}

In Table~\ref{lstab:sfxt} an updated list of SFXTs is reported with some interesting quantities 
(spectral type of the supergiant companion, 
source distance, orbital geometry, spin period and super-orbital periodicities).
Some of these properties are apparently very similar to those displayed by some persistent SgHMXBs:    
the spectral types of the massive companion overlap in the two kinds of HMXBs (Fig.~\ref{lsfig:spetype}, 
although O-type donors might appear more popular among SFXTs 
than in classical SgHMXBs);
SgHMXBs have orbital periods overplapping with those shown by SFXTs, as well as pulsar spin periods  (Fig.~\ref{lsfig:corbet}),
although SFXTs orbital periodicities span a wider range of values. 

X--ray pulsars appear to be more elusive among SFXTs, despite extensive searches; this might be due either to the 
large X--ray variability during flaring activity 
(hampering the detection of pulsations) or to the fact that some spin periods might be very long. 
The spin period in IGR~J17544-2619 is controversial (marked by a question mark in Fig.~\ref{lsfig:corbet}),
since it has been found with $RXTE$/PCA data, and might be due  to an another transient source within 
the field of view \cite{Drave2012}. 
Alternatively, the lack of pulsations in many SFXTs could indicate that the compact object is a black hole,
but the similarity of their X-ray spectra with those displayed by accreting X--ray pulsars 
(flat power-law within 10 keV, with a cutoff around 10--30~keV)
suggests to exclude this possibility.

\begin{table*}
\setlength{\tabcolsep}{2pt}
 \centering
  \caption{Properties of SFXTs.}\label{lstab:sfxt}
\begin{footnotesize}
  \begin{tabular}{@{}lllllll@{}}
\hline
   Name & Companion & Distance & Orbital          &  ecc  &   Spin            &  Super-Orbital     \\
        &           &  (kpc)     &  Period (d)      &                &  Period (s)       &  Period (d)       \\
\hline
IGR~J08408--4503  & O8.5Ib$^{\cite{Barba2006:08408-4503}}$                                 &   2.7$^{\cite{Leyder2007}}$                     & 9.5436$^{\cite{Gamen2015}}$                                       & 0.63$^{\cite{Gamen2015}}$      &     ---    & 285 $\pm$ 10$^{\cite{Gamen2015}}$  \\
IGR~J11215--5952  &  B0.5Ia$^{\cite{Negueruela2005hd}, \cite{Lorenzo2010}, \cite{Lorenzo2014}}$   &  $>$7$^{\cite{Lorenzo2014}}$             &  164.6$^{,\cite{Sidoli2006, Sidoli2007, Romano2009:11215_2008}}$  & $>$0.8$^{\cite{Lorenzo2014}}$  &  187$^{\cite{Swank2007, Sidoli2017}}$   & ---                     \\
IGR J16328--4726  & O8Iafpe$^{\cite{Hanson1996}, \cite{Coleiro2013}, \cite{Fiocchi2010}}$  &  7.2 $^{\cite{Persi2015}}$                      &  10.068$^{\cite{Corbet2010}, \cite{Fiocchi2013}}$ &   ---     &   ---                          & ---  \\
IGR~J16418--4532  &  OB Sg                                                                 &  $\sim$13$^{\cite{Chaty2008}}$                  &  3.753$^{\cite{Corbet2006}}$                       &   ---    & 1212$^{\cite{Sidoli2012b}}$    & 14.6842$^{\cite{Corbet2013}, \cite{Drave2013Atel}}$  \\ 
IGR~J16465--4507  & O9.5Ia$^{\cite{Coe1996}}$                                              & 9.5$^{\cite{Clark2010}}$                        & 30.243$^{\cite{LaParola2010}, \cite{Walter2006}}$  &   ---    & 228$^{\cite{Lutovinov2005igr16465}}$   & --- \\  
IGR~J16479--4514 & O8.5I, O9.5Iab$^{\cite{Rahoui2008}, \cite{Nespoli2008}}$                & 4.9, 2.8$^{\cite{Rahoui2008}, \cite{Nespoli2008}}$  & 3.3194$^{\cite{Jain2009}, \cite{Romano2009}}$  &   ---    &     ---                        &   11.880$^{\cite{Corbet2013}, \cite{Drave2013Atel}}$  \\
XTE~J1739--302   & O8Iab(f)$^{\cite{Negueruela2006}, \cite{Rahoui2008}}$                   & 2.7$^{\cite{Rahoui2008}}$                       & 51.47$^{\cite{Drave2010}}$                        &   ---    &     ---                        &  ---          \\
IGR~J17544--2619 & O9Ib$^{\cite{Gimenez2016}}$                                             & 3.0$^{\cite{Gimenez2016}}$                      & 4.926$^{\cite{Clark2009}}$                        &    $<$0.4$^{\cite{Clark2009}}$  &  71.49$^{\cite{Drave2012}}$  & ---      \\
IGR~J17354--3255 &   OB Sg(?)                                                              &  8.5$^{\cite{Tomsick2009b}}$                    &  8.4474$^{\cite{Dai2011}, \cite{Sguera2011}}$     &   ---    &         ---                  & ---      \\
SAX~J1818.6--1703 & $\sim$B0I$^{\cite{Negueruela2006:aTel831}, \cite{Torrejon2010}}$       & 2, 2.1$^{\cite{Negueruela2008int}, \cite{Zurita2009}}$ & 30.0$^{\cite{Bird2009}, \cite{Jain2009}}$   &   --- &         ---           &  ---     \\
IGR~J18410--0535 &  B1Ib$^{\cite{Nespoli2007:18410-0535}}$                                 &   3.2$^{\cite{Walter2006}}$                     &    ---                                             &   --- &         ---           & ---  \\
IGR~J18450--0435 & O9.5I$^{\cite{Coe1996}}$                                                & 3.6$^{\cite{Coe1996}}$                          &  5.7195$^{\cite{Goossens2013}}$                    &   --- &         ---           &  ---  \\
IGR~J18462--0223 &   OB Sg(?)                                                              & 11(?)$^{\cite{Sguera2013}}$                     & 2.13(?)$^{\cite{Sguera2013}}$                      &   --- &         ---           &  ---  \\
IGR~J18483--0311 &  B0.5Ia$^{\cite{RahouiChaty2008}}$                                      & 3$^{\cite{RahouiChaty2008}}$                    &  18.55$^{\cite{Levine2006}}$                       &   $\sim$0.4$^{\cite{Romano2010igr18483}}$ &  21.0526$^{\cite{Sguera2007}}$   &   --- \\
\hline
\multicolumn{6}{@{}p{0.95\textwidth}}{\vspace*{0.2\baselineskip}%
}
\end{tabular}
\end{footnotesize}
\end{table*} 

\section{Proposed explanations for the SFXT behaviour}

The whole core debate on SFXTs revolves around the physical mechanism producing their X--ray phenomenology,
at odds with what is observed in persistent SgHMXBs: 
their low percentage of time spent in bright X--ray flares, 
their large X--ray intensity variability between quiescence and flare peaks, 
their low time-averaged X--ray emission.

At first, it was suggested that the sporadic flares are  
produced by the direct (Bondi-Hoyle) accretion of supergiant wind clumps onto the neutron star (\cite{zand2005}).
But in this case, the estimated mass of the clumps can reach huge 
values ($10^{21}$~--$10^{23}$~g), too large compared with what is known 
about hot stellar winds (clump masses $\lsim$$10^{18}$~g; \cite{Martinez-Nunez2017}).
It was demonstrated by \cite{Oskinova2012} that the direct accretion from clumpy winds 
cannot account for SFXTs light curves, thus 
requiring a mechanism able to reduce  accretion rate most of the time
and to produce the rare and luminous  X--ray flaring activity with the needed dynamic range.

Nowadays, the most debated explanations for the SFXT behaviour are 
the centrifugal (or magnetic) inhibition of accretion (``gating'' mechanisms)
at the neutron star magnetosphere (\cite{Grebenev2007}, \cite{Bozzo2008}) 
and the subsonic settling accretion regime (\cite{Shakura2012}, \cite{Shakura2017}, \cite{Shakura2014}, \cite{Postnov2017}).

Gating mechanisms critically depend on the pulsar spin period and the neutron star magnetic field (see \cite{Bozzo2008} for a review),
in presence of a variability in the density and velocity of the clumpy wind at the neutron star orbit.
The centrifugal barrier can be at work in neutron stars with standard magnetic fields (B$\sim$10$^{12}$~G) 
and pulsar spin periods of $\sim$10~s, assuming blue supergiants winds with typical parameters (terminal velocity $\sim$1000~km~s$^{-1}$, 
mass-loss rate $\mdot$$\sim$10$^{-6}$~$\msun$~yr$^{-1}$). Of course, this also depends on the orbital geometry. 
In presence of longer  spin periods, the neutron star magnetic field should be higher, for the centrifugal barrier to be closed.
The magnetic barrier operates in slow pulsars (spin periods exceeding 1000~s) with magnetar-like magnetic fields  (B$\sim$10$^{14}$~G).
These mechanisms prevent accretion onto the neutron star most of the time, but the encounter with a dense clump 
along the orbit might temporarily allow the accretion, triggering  an X--ray flare.  
Gating mechanisms require that SFXTs host slow spin periods ($>$1000~s) and very strong magnetic fields, 
to reproduce their light curves \cite{Bozzo2016gating}, implying that 
SFXTs should host slower pulsars with much higher magnetic fields than in persistent SgHMXBs.

\begin{figure}[ht!]
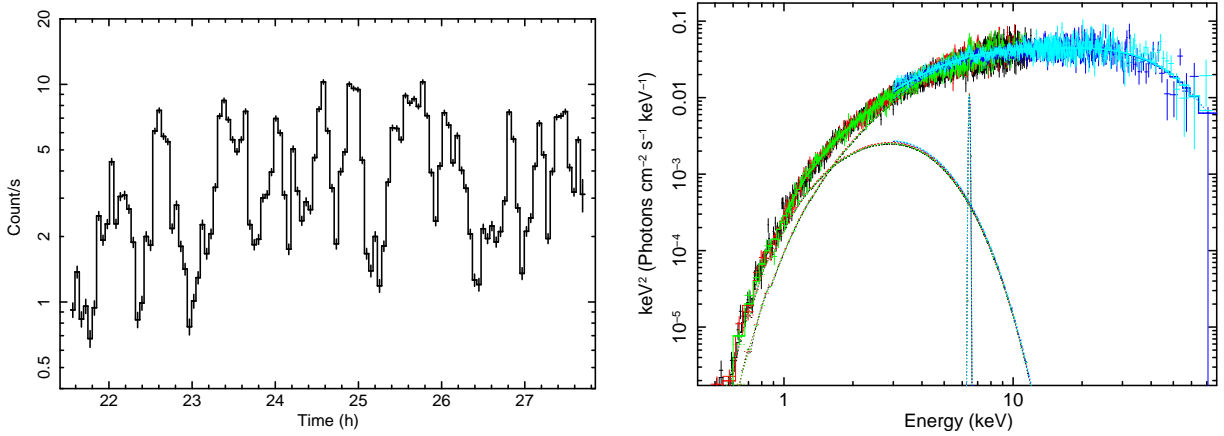

\begin{center}
\begin{tabular}{cc}
\hspace{-0.9cm}
\includegraphics*[angle=-90,scale=0.33]{lsfig4a_new.ps} 
\includegraphics*[angle=-90,scale=0.32]{lsfig4b_new.ps} \\
\end{tabular}
\end{center}
\caption{\scriptsize {\em Left panel:} EPIC pn light curve of SFXT IGRJ~11215-5952 during 
the February 2016 periodic outburst (bin time of 187~s, the pulsar spin period).
{\em Right panel:} Broad-band time-averaged spectrum of the SFXT IGRJ~11215-5952 
observed by \xmm\ and $NuSTAR$ in February 2016, fitted with a soft black body (kT$_{bb}$=0.7~keV and a radius of few hundred meters at 7~kpc), 
a Comptonized emission (electron temperature
kT$_{e}$$\sim$9~keV) and a narrow emission line by neutral iron at 6.4~keV (\cite{Sidoli2017}).
}
\label{lsfig:11215}
\end{figure}

Neutron star magnetic fields (directly measured from the detection of cyclotron resonance scattering features in the X--ray spectrum) 
determined to-date in SFXTs are the following:  
a hint of a low magnetic field (B$\sim$10$^{11}$~G) has been obtained in IGR~J18483--0311 (\cite{Sguera2010}), while
a few 10$^{12}$~G has been measured in IGR~J17544--2619 from the detection of a cyclotron line at $\sim$17~keV 
in the $NuSTAR$ spectrum \cite{Bhalerao2015}. However the presence of this cyclotron line was not confirmed by \cite{Bozzo2016} 
during a second $NuSTAR$ observation that caught the source at similar X--ray fluxes, possibly suggesting a variability of the cyclotron line.
Note also that \cite{Bozzo2016} failed to detect any pulsar period from IGR~J17544--2619.
A search for a cyclotron line in the \xmm\ plus $NuSTAR$ X--ray spectrum of the SFXT pulsar IGR~J11215--5952  \cite{Sidoli2017}
resulted in a hint of an absorption feature at 17 keV during a couple of flares. With an estimated line significance at 2.63~$\sigma$, 
this line needs further confirmation 
(see Fig.~\ref{lsfig:11215} for the source light curve in outburst, and the time-averaged spectrum during the observation 
performed in February 2016).

An alternative model proposed to explain the SFXT behaviour is
the settling accretion regime (\cite{Shakura2012}, \cite{Shakura2017}) that operates in slow, low luminosity,
HMXB pulsars (L$_{X}$$<$4$\times$10$^{36}$~erg~s$^{-1}$).
Below this luminosity, the matter captured within the Bondi radius
is unable to efficiently cool down (by Compton processes) to penetrate the neutron star
magnetosphere by means of Rayleigh-Taylor instabilities. 
A quasi-spherical shell of hot matter forms above the magnetosphere, 
and only a reduced accretion rate (with respect to what expected in Bondi direct accretion, that operates at higher L$_{X}$) 
is permitted, by means of inefficient radiative cooling. 
The low time-averaged luminosity observed in SFXTs suggests that settling accretion can set-in in these sources. 
In this scenario, the sporadic bright X--ray flares are produced by the complete 
collapse of the shell onto the neutron star.
The accreted mass needed to fuel the X--ray flare well agrees with the estimated
mass of the quasi-spherical shell (\cite{Shakura2014}), without the need for huge wind clumps.
Also a correlation between the energy emitted during SFXTs flares and the low time-averaged luminosity
has been observed, as predicted by this model (\cite{Shakura2014}).

The trigger for the sporadic instability of the shell has been suggested to reside in the magnetized stellar wind from the
supergiant donor: magnetic reconnection can temporarily open the neutron star magnetic field lines, allowing accretion
of the accumulated matter on free-fall time scale (\cite{Shakura2014}). 
%


Cumulative distributions of X--ray flare luminosities  observed with \inte\ (17--50 keV) 
support the settling accretion model (\cite{Paizis2014}): 
the exploitation of the whole \inte\ archive spanning more than 10 years, allowed us to build
meaningful distributions of SFXTs peak flare luminosities.
The main results of this systematic study were the full characterization of the SFXT low duty cycles in outburst (0.1-5$\%$), 
together with the important findings that the 
luminosity distributions of the SFXTs flares follow a power-law, contrary
to what displayed by SgHMXBs, where the distribution of the X--ray luminosity is log-normal.
Power-law-like distributions 
are reminiscent of self-organized criticality (SOC) systems (see the reviews by \cite{Aschwanden2013}, \cite{Aschwanden2017}).
The prototypical example of a SOC system is a sandpile, where unpredictable sand avalanches occur when
an instability threshold is reached by adding sand grains, one by one, to the pile; when a critical
slope of the pile is reached,  sand avalanches occur and  they display many sizes. The resulting size distribution is a power-law (\cite{Bak1987}).
Also solar flares, among other natural phenomena, are thought to be SOC systems and produce power-law-like size distributions.
Since solar flares are produced by magnetic reconnection, SFXTs flares (produced by the collapse of the shell) 
could be triggered by the interaction of the accreting magnetized stellar wind with the neutron star magnetosphere (\cite{Shakura2014}).
Also the temporal properties of flares from SFXTs show cumulative distributions that can be interpreted in the scenario 
that involves magnetized supergiant winds and their fractal structure \cite{Sidoli2016}.

As the magnetic field of the accreting wind material is proposed to play an important role 
in triggering the flares by magnetic reconnection at the neutron star magnetospheric boundary, 
also the density and velocity of the clumpy wind  have  been proposed as 
a crucial difference with respect to persistent SgHMXBs (\cite{Shakura2014}, \cite{Postnov2017}).
Systematically faster and less dense winds in massive donor in SFXTs could 
reduce the X--ray luminosity in order to enable the inefficient radiative cooling regime in the quasi-spherical settling accretion. 

It is usually assumed that supergiants in SFXTs and in persistent SgHMXBs are very similar. 
However, this paradigm needs to be tested observationally, and deserves further multifrequency investigation.
Very recently, a comparative study of the wind in the donor stars in  IGR J17544-2619 (the prototype of SFXTs) 
and Vela X-1 (the prototype of persistent SgHMXBs) revealed that these supergiants have different properties:
the wind terminal velocity is observed to be a factor of two higher in this SFXT (1500~km~s$^{-1}$) 
than in Vela X-1 (700~km~s$^{-1}$, \cite{Gimenez2016}). 
A faster wind implies a much lower accretion rate in IGR J17544-2619, with respect to Vela X--1.

These important observations strongly support the idea that the dichotomy of wind-fed HMXBs, between SFXTs and persistent sources,
is produced by a dichotomy in the supergiant wind properties \cite{Gimenez2016}.
However, further multifrequency observations of members of both classes are needed to provide us with 
a final answer to the issue, since the X--ray behavior emerges from the specific combination
of the properties of both the neutron star (spin and magnetic field) and the supergiant wind (its velocity, density and magnetic field).

\section{Conclusions and future directions}

In this brief review paper I have summarized the most recent achievements in the SFXT field.
The two main proposed explanations for the SFXTs phenomenology (gated mechanisms and 
subsonic settling accretion regime) invoke a dichotomy in the
neutron star properties and/or in the wind from the hot massive companion, between SFXTs and persistent 
SgHMXBs.
Neutron star properties in SFXTs remain quite elusive and require new X--ray observations.
Given the power-law distribution of the SFXTs flare luminosities, the future discovery of
more extreme behaviours at X--rays is certain: it is only a matter of (observing) time.

HMXBs are not clean laboratories and the donor and the neutron star
are inextricably intertwined: a complex feedback is present between the two protagonists,
since also the X--ray emission from the compact object influences the wind structure and its acceleration  
by means of photoionization (\cite{Martinez-Nunez2017}).
For this reason, systematic, simultaneous, orbital-phased resolved, multifrequency observations 
would be desired to solve the issue of the SFXT variability.
In particular, after more than a decade from their discovery, a  dedicated effort 
to the investigation of the wind of their supergiant donors is urgently needed.

\acknowledgments

I would like to thank the organizers for their kind invitation to give this review talk 
at the {\em XII Multifrequency Behaviour of High 
Energy Cosmic Sources Workshop}, held in Palermo, Italy, in June 2017.

I acknowledge financial contribution from the agreement ASI-INAF INTEGRAL 
 n. 2013-025.R.0, from the agreement ASI-INAF NuSTAR I/037/12/0,
and from the PRIN-INAF~2014 grant 
``Towards a unified picture of accretion in High Mass X-Ray Binaries''. 



\end{document}